\documentclass[11pt, a4paper]{article}

\usepackage[utf8]{inputenc} 
\usepackage[T1]{fontenc}    
\usepackage{lmodern}        
\usepackage[a4paper, top=2.5cm, bottom=2.5cm, left=2.5cm, right=2.5cm]{geometry}

\usepackage{amsmath}
\usepackage{amssymb}
\usepackage{graphicx}
\usepackage{titlesec}
\usepackage{authblk}
\usepackage{url}
\usepackage{hyperref}
\usepackage{setspace}
\usepackage{caption}
\usepackage{subcaption}
\usepackage[numbers]{natbib}

\usepackage{pgfplots}
\pgfplotsset{compat=1.18}
\usepackage{tikz}

\usepackage{float}

\title{\textbf{Low Energy Neutrinos in Milky Way and  Cloud-9}}

\author{Eduardo Flores\footnote{flores@rowan.edu}}
\author{Elise Cantu}
\author{Ian Marano}
\author{Osvan Vivar-Garcia}
\author{Shabhaz Khalandar}
\author{Patrick Walker}

\affil{Department of Physics \& Astronomy, Rowan University}

\date{November 24, 2025}

\begin{document}

\maketitle

\begin{abstract}

We study low-energy galactic neutrinos in the Milky Way under two fundamentally different descriptions of gravity, showing that neutrinos provide a sensitive probe of gravity’s underlying nature. If gravity is a quantum interaction, its long-range character leads to the formation of an atom-like bound neutrino structure. We compute its mass distribution and find that, within a radius  $292kpc$, the total mass is only  $10^{-29}$ of the galaxy’s dark matter, ruling it out as a dark-matter candidate. Nevertheless, experimental confirmation of this structure would constitute direct evidence for gravity as a quantum force mediated by gravitons. If gravity instead arises from spacetime curvature, neutrinos interact only through the short-range weak force and are therefore effectively collisionless. In this regime, neutrinos behave as free classical particles orbiting the galaxy and do not experience Fermi pressure. We show that such a population can be sufficiently compact to reproduce the Milky Way rotation curve, thereby making neutrinos viable dark-matter candidates within this framework. We further model Cloud-9’s dark-matter component as a population of free, low-energy, collisionless neutrinos. Because the neutrino–antineutrino annihilation cross section is extremely small, neutrinos and antineutrinos may remain in near equilibrium over cosmological timescales, potentially providing a mechanism relevant to the observed matter–antimatter asymmetry. 

\end{abstract}

\newpage

Low-energy galactic neutrinos gravitationally bound to the galaxy are nearly ghost-like particles. Their interactions with matter are practically undetectable due to their extremely low energies, small masses $m\leq0.4 eV/c^2$ \cite{KATRIN2025}, and the very short range of the weak interaction, approximately $10^{-18}m$. In fact, there could be as many—or even more—antineutrinos of comparable energy, and they would remain equally undetectable. Neutrino–antineutrino pairs experience mutual attraction and can annihilate only into photons; however, the cross section for this process is extremely small \cite{degraaf1966}. As a result, very low-energy neutrinos and antineutrinos would remain in equilibrium. It is therefore conceivable that a sufficiently large population of antineutrinos could exist to restore particle–antiparticle symmetry in the universe. Nevertheless, using the Vlasov equation, Tremaine and Gunn showed that massive galactic halos cannot be composed of stable neutral leptons with masses below $1 MeV/c^2$ \cite{tremaine1979}.

Despite this constraint, recent developments concerning the role of virtual particles in particle interactions motivate a reassessment of light neutrinos as potential dark-matter candidates. In particular, Flores \cite{flores_qed} argues that when virtual particles are absent—for whatever reason—no physical interaction can occur. Moreover, without virtual particles, particles cannot follow the prescriptions of their associated wavefunctions and instead behave as free classical objects bound only by conservation laws. This is also the case for Fermi pressure, which arises from the symmetry properties of the wavefunction. Fermi pressure is a physical effect that can be of significant magnitude and therefore requires energy and momentum exchange among identical fermions—exchanges that must be mediated by virtual particles. We therefore conclude that, in the absence of virtual particles, Fermi pressure cannot arise.

In quantum field theory, interactions between particles are mediated by the exchange of virtual particles \cite{bjorken1964}. Virtual particles exhibit unusual properties: they cannot be isolated, and they exist only for extremely short times. A key characteristic of a virtual particle is the interaction range it mediates. For example, the interaction range of a virtual photon or virtual graviton is, in principle, infinite, whereas the range associated with a virtual massive vector boson is on the order of  $10^{-18}$ meters. Applying these considerations to neutrinos leads to two distinct possibilities, depending on the nature of gravity. If gravity is a quantum interaction, virtual gravitons exist and neutrinos experience a long-range interaction mediated by them. If, instead, gravity is purely a manifestation of spacetime curvature, as described by general relativity, neutrinos interact only through the weak force, whose extremely short range renders low-energy neutrinos effectively collisionless. In this case, low-energy galactic neutrinos would behave as a classical gas.

We first consider that gravity is a quantum force, then neutrinos have a long range interaction mediated by virtual gravitons. Let us assume that non-relativistic neutrinos of mass $m=0.4 eV/c^2$ are the dominant component of dark matter, we calculate some of its properties. The dark-matter density near the solar system is estimated to be $10^{-21}kg/m^3$ \cite{Xu}. The corresponding number density is $n=1.4\times10^{15} neutrinos/m^3$. The temperature to keep these neutrinos bound to the galaxy at this location is of the order of  $T=5.2$ m$K$. Then, the degeneracy term is extremely large 
\begin{equation}
    \frac{1}{2^{5/2}} \frac{n h^2}{(2\pi mkT)^{3/2}}=4\times10^5.
    \label{eq:degeneracy}
\end{equation}
This neutrino gas is a completely degenerate Fermi gas since the value in Eq.  \eqref{eq:degeneracy} is much larger than $1$. These neutrinos experience Fermi pressure. In fact, these neutrinos behave as fermions at zero Kelvin. 

Neutrinos stronger interaction with each other is via weak interaction, with an effective range, $d=10^{-18}m.$ To estimate how often neutrinos interact with each other via weak interaction let us treat them as a classical gas and compute their mean free path, 
\begin{equation}
    \lambda=\frac{1}{\sqrt{2}\pi d^2n}=1.6\times10^{20}m\approx 300 kpc.
   \label{eq:mean_free_path}
\end{equation}
Thus, a neutrino can easily travel across the galaxy and not interact with another neutrino. The fact that low energy galactic neutrinos are quite free suggests that these neutrinos would tend to form a large neutrino atom.

The neutrino atom model posits that the galactic halo is a macroscopic quantum bound state, where neutrinos occupy discrete quantum levels defined by the gravitational potential of the galaxy. Just as the stability of the Hydrogen atom relies on the exchange of virtual photons \cite{greiner2009}, as described by QED, the stability of the Galactic Neutrino Atom relies on the exchange of virtual gravitons \cite{flores_qed}.

The hydrogen atom is typically studied by solving the Dirac equation applicable to QED. We assume that galactic neutrinos follow the Dirac equation in Schwarzschild spacetime \cite{Cotaescu}. The solution to the angular part of this Dirac equation is similar to the solution of the Dirac equation in Hydrogen. The radial part is complicated and can only be solved as an approximation. Fortunately, we only need to analyze the solutions far from the singularity for which we have good approximations \cite{Cotaescu}. The discrete quantum modes with discrete energy levels, $\epsilon<\mu$,  are governed by radial wave functions which must be square integrable. Note that here $\mu$, is proportional to the neutrino mass, $\mu=mR$, where $R$ is the Schwarzschild radius. The  solutions are given by
\begin{equation}
    f^ +_a (x) = N ^+x^{2s^j} {e^{-\nu x^2}}_1 F_1(s_j+q+1,2s_j+1,2\nu x^2),
    \label{eq:wavefunction+}
\end{equation}
\begin{equation}
f^-_a(x)=N^-x^{2s^j} {e^{-\nu x^2}}_1 F_1(s_j+q,2s_j+1,2\nu x^2),
\label{eq:wavefunction-}
\end{equation}
where $x\approx\sqrt{r/R}$ and the confluent hypergeometric functions ${}_1F_1$ depend on the following real parameters
\begin{equation}
    s_j =\sqrt{\kappa^ 2_ j + \nu ^2 + \mu^2 (\delta^ 2 -1)},
\end{equation}
\begin{equation}  
    q = \nu +\frac{2\mu  (\delta-1)}{\nu} .
\end{equation}
The normalization constants $N^+$ and $N^-$ must obey
\begin{equation}
    \frac{N^-}{N^+}=\frac{\nu(s_j-q)}{\kappa_j\nu-\mu\epsilon(\delta-1)}.
\end{equation}
These solutions may be square integrable with respect to the scalar product only when we imposed the quantization condition $s_j+q=-n$,  $n=1,2,3,...$  This leads to the equation
\begin{equation}
    \sqrt{\kappa^ 2_ j + \nu ^2 + \mu^2(\delta^ 2-1)}+\nu +\frac{\mu^2(\delta-1)}{\nu}=-n,
\end{equation}
which is of the third order in $\nu$ and can be easily solved in Mathematica or equivalent system. Once we obtain $\nu$,  we can plot the radial probability distribution of the corresponding wavefunctions in Eqs.  \eqref{eq:wavefunction+} and \eqref{eq:wavefunction-}. 

\begin{figure}[H]
    \centering
    \begin{subfigure}[b]{0.48\textwidth}
        \includegraphics[width=\textwidth]{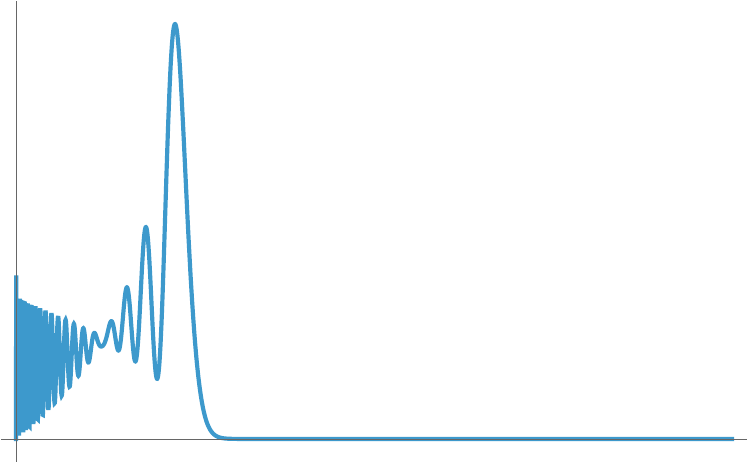}
        \caption{$n=21,\kappa=1$}
        \end{subfigure}
    \hfill
    \begin{subfigure}[b]{0.48\textwidth}
        \includegraphics[width=\textwidth]{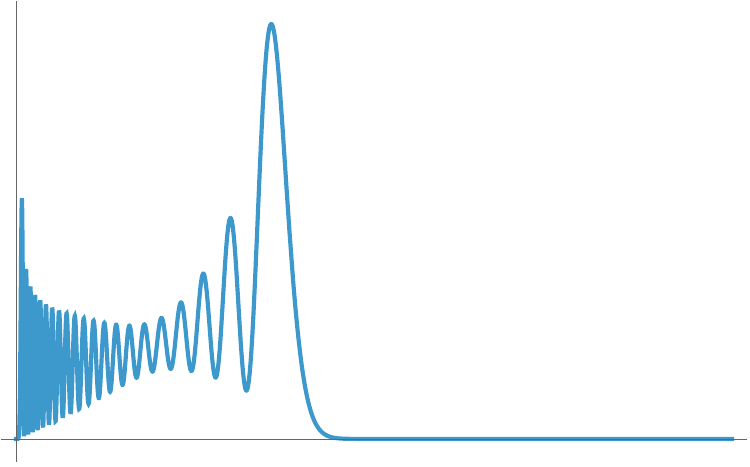}
        \caption{$n=1,\kappa=7$}
        \end{subfigure}
    \vfill
    \begin{subfigure}[b]{0.48\textwidth}
        \includegraphics[width=\textwidth]{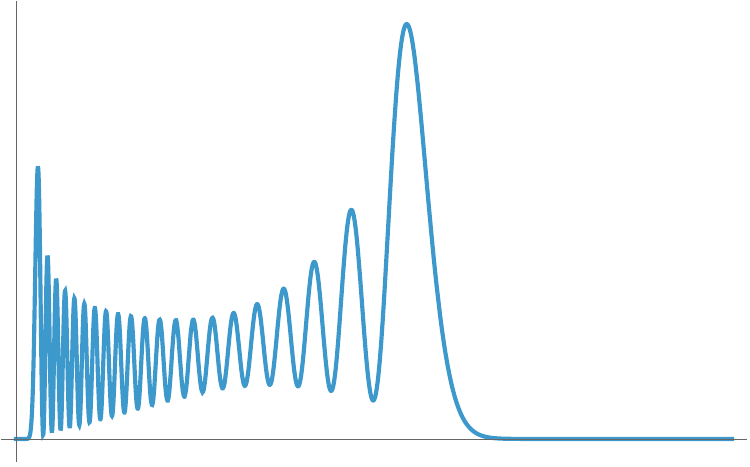}
        \caption{$n=21,\kappa=14$}
        \end{subfigure}
    \hfill
    \begin{subfigure}[b]{0.48\textwidth}
        \includegraphics[width=\textwidth]{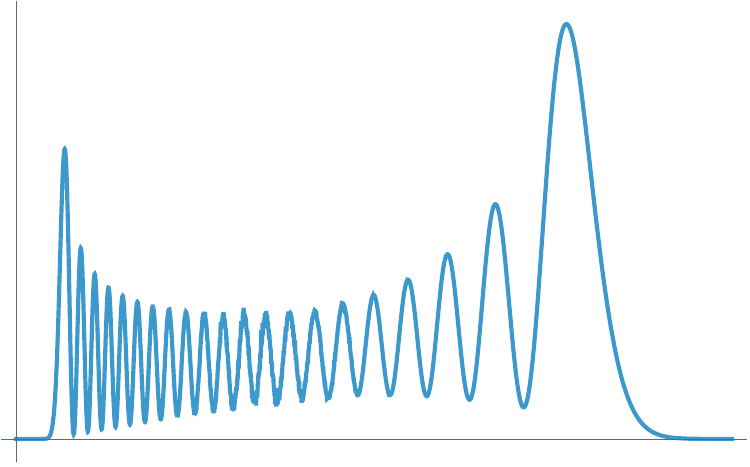}
        \caption{$n=21,\kappa=21$}
        \end{subfigure}
    \caption{\textbf{Probability distributions for single particle in n=21.} We plot here few probability distributions for principal quantum number n=21 and a sample of angular momenta. We note that all the distributions from (a) to (d) have similar shape and only differ on how stretched they are. We note that all the distributions have 21 peaks. In all cases, the plot range starts at $10R$.}
  \label{fig:prob_dist}
\end{figure}

Our purpose is to find the radial probability distribution of the neutrino atom with as many as possible particles filling all the lowest quantum states. In Fig. \ref{fig:prob_dist} we plot few probability distribution obtained from wavefunctions with similar principal quantum number but different angular momentum. We find that the wavefunctions in the Schwarzschild metric and the hydrogen atom are both described by the confluent hypergeometric functions ${}_1F_1$. However, there are major differences. For instance, in the Schwarzchild metric, the distribution that corresponds to  $n=21$ and $\kappa=21$, has 21 peaks, while in the Hydrogen atom, the corresponding distribution has only 1 peak.

\begin{figure}[H]
    \centering
    \begin{subfigure}[b]{0.48\textwidth}
        \includegraphics[width=\textwidth]{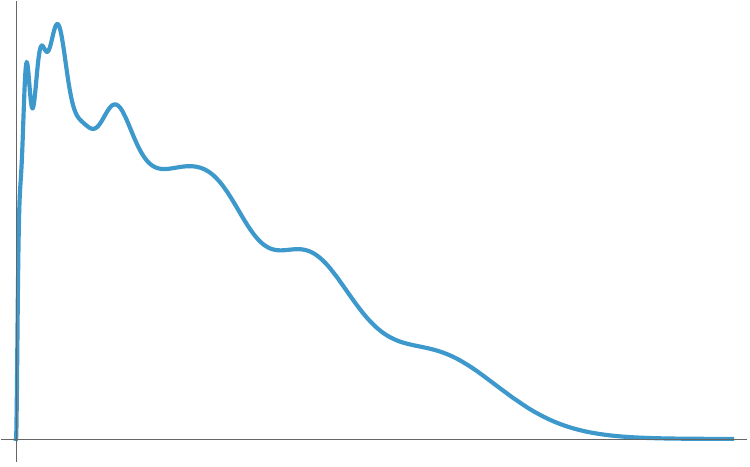}
        \caption{$n=6,  N=72$}
              \label{fig:particles_dist}
    \end{subfigure}
    \hfill
    \begin{subfigure}[b]{0.48\textwidth}
        \includegraphics[width=\textwidth]{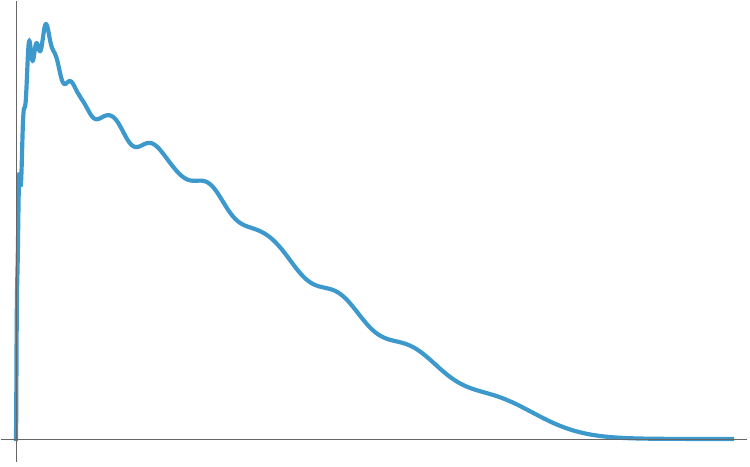}
        \caption{$n=11, N=241$}
    \end{subfigure}
    \vfill
    \begin{subfigure}[b]{0.48\textwidth}
        \includegraphics[width=\textwidth]{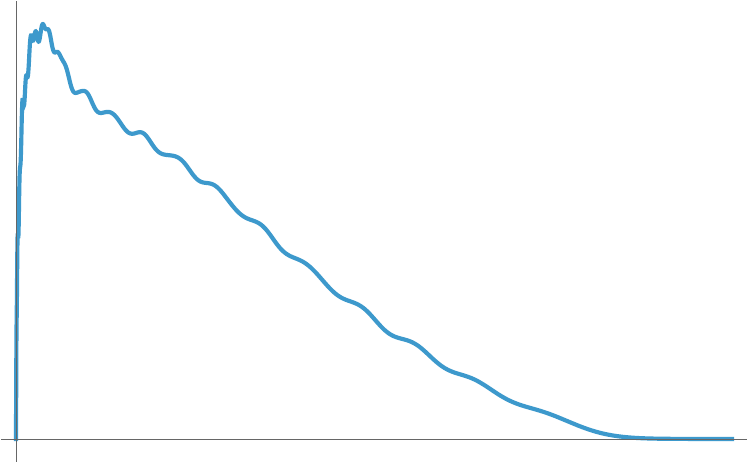}
        \caption{$n=16,  N=512$}
    \end{subfigure}
    \hfill
    \begin{subfigure}[b]{0.48\textwidth}
        \includegraphics[width=\textwidth]{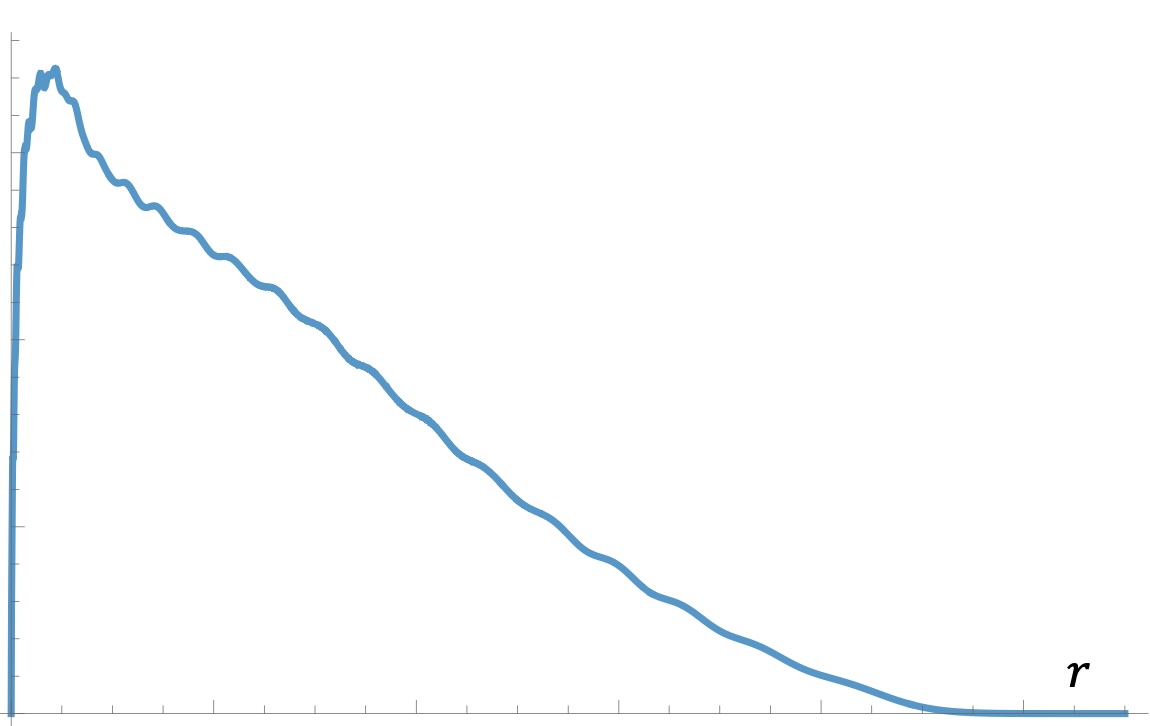}
        \caption{$n=21,  N=882$}
    \end{subfigure}
    \caption{\textbf{Probability distributions for many particles at once.} We plot few probability distributions for principal quantum number ranging from (a) $n=6$ composed of $N=72$ normalized wavefunctions to (d) $n=21$ composed of $N=882$ normalized wavefunctions. We notice that the triangular shape of the distribution get more accentuated with higher principal number $n$.  In all cases, the plot range starts at $10R$.}
    \label{fig:particles2_dist}
\end{figure}

The important thing for us is to determine the radial probability distribution of a large number of neutrinos. The wavefunctions from the Dirac equation in Schwarzschild metric are not as easy to handle as the related wavefunctions for the hydrogen atom. As the principal quantum number $n$ increases, the associated wavefunctions of the Dirac equation in Schwarzschild metric are highly oscillatory, making their plots unreliable. Thus, we determine the probability distribution pattern by plotting a number of normalized wavefunctions with low quantum number $n$. We want to fill all the lowest states starting with $n=1$. In Fig. \ref{fig:particles_dist}, we fill every state up to $n=6$ and notice a triangular shape of the distribution made up of 72 wavefunctions. As we continue to higher $n$ we see that the shape of a triangle is more obvious.

\begin{figure}[H]
    \centering
    \begin{subfigure}[b]{0.48\textwidth}
        \centering

\includegraphics[width=\textwidth]{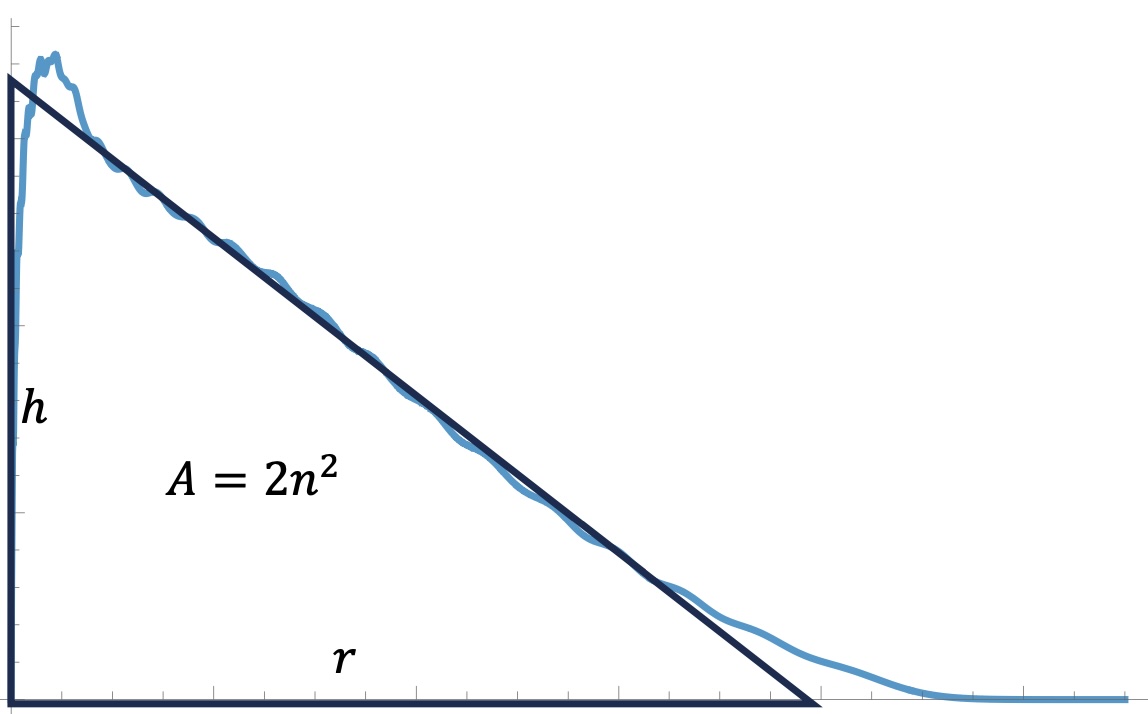}
    \end{subfigure}
    \begin{subfigure}[b]{0.48\textwidth}
   
    \end{subfigure}
    \caption{\textbf{Triangular fit.} We see that as the number of particles in a distribution that starts at $n=1$ and fills every state to a given $n$ increases, the shape approaches a right triangle. We assume that the triangular shape is maintained as $n$ grows large. The area of the triangle represents the number of particles $A=N=2n^2$. The size of the base, $r$, is 4/3 of location of the peak of the asymptotic wavefunction with largest $n$ and $\kappa=n$.  The height, $h$, is defined by $A$ and $r$.}
    \label{fig:tria_fit}
\end{figure}

We observe that as the number of wavefunctions, $N$, that make up the distribution increases, the shape of the distribution approaches a right triangle as seen in Fig. \ref{fig:tria_fit}. We exploit this feature to generalize the distribution as a function of $r$ and $n$.

\begin{figure}[H]
    \centering
    \begin{subfigure}[b]{0.48\textwidth}
        \includegraphics[width=\textwidth]{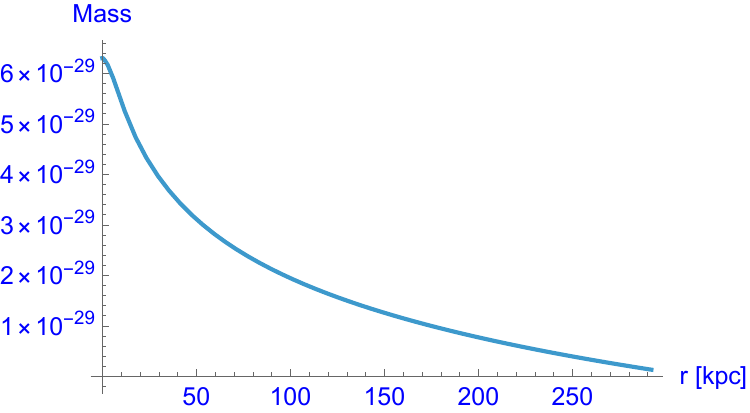}
        \caption{Neutrino atom as dark matter}
              \label{fig:qua_fit}
    \end{subfigure}
    \hfill
    \begin{subfigure}[b]{0.48\textwidth}
        \includegraphics[width=\textwidth]{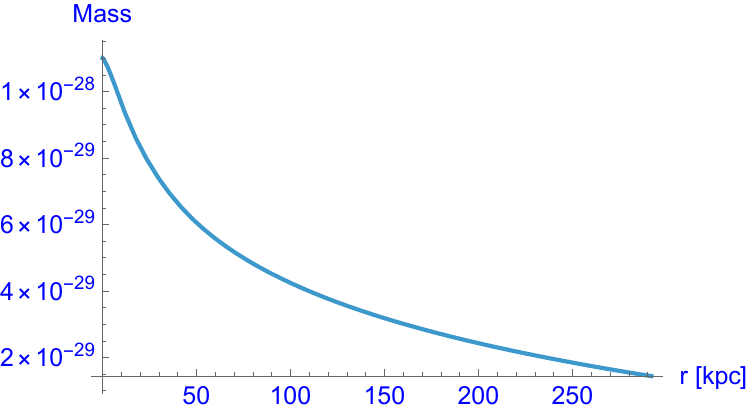}
        \caption{Neutrino atom in the presence of dark matter}
        \label{fig:qua2_fit}
    \end{subfigure}

    \caption{\textbf{Mass probability distribution for the neutrino atom.} The units are in $10^9\textup{M}_\odot/kpc$. This distribution ranges from nearly the center of the galaxy to nearly the edge at 292 kpc.   (a) Assuming that the neutrino atom is dark matter we obtain a mass distribution made of neutrinos. The area under this graph is of order of $10^{-29}$ times smaller that the expected dark matter in the Milky-Way galaxy. (b) The neutrino atom is slightly more massive in the presence of dark matter not made of neutrinos.}
\end{figure}

From Fig. \ref{fig:tria_fit} we obtain the slope of the probability distribution,
\begin{equation}
    \text{Slope} =-\frac{h}{r}= -\frac{4n^2}{r^2} .
    \label{eq:slope}
\end{equation}
We obtained $h$ from the area, $A=\frac{1}{2}rh=2n^2$. To find how $r$ is a function of $n$, we use an asymptotic approximation, for large argument, alternative to the confluent hypergeometric function, ${}_1F_1(a,b,z)$, known as $U(a,b,z) \approx z^{-a}$ \cite{LDMF}. Inserting this new function into Eqs.  \eqref{eq:wavefunction+} and \eqref{eq:wavefunction-} and using the largest principal quantum number $a=-n$ and $\kappa=n$, we find a single peak located at 
\begin{equation}
    \frac{4n^2 \hbar^2}{G m^2M_{total}},
    \label{eq:peak}
\end{equation}
where $M_{total}$ is the total mass within the atom. The location of this peak is about 3/4 of the base, $r$, of the triangle in Fig. \ref{fig:tria_fit}.  Thus, we estimate the base, $r$, of the triangle to be
\begin{equation}
    r =\frac{4}{3} \frac{4n^2 \hbar^2}{G m^2( M+2n^2m)},
    \label{eq:base}
\end{equation}
where $( M+2n^2m)$ is the total mass, $M_{total}$, contained within radius $r$, and $M$ is baryonic mass. We solve Eq. \eqref{eq:base} for $n$, replace it in the slope in Eq. \eqref{eq:slope} and we have
\begin{equation}
    \text{Slope}=-\frac{6 G m^2 M(r)}{(-3 G m^3 r + 8\hbar^2)r }.
    \label{eq:Slope2}
\end{equation}
To obtain the mass probability distribution, up to constant of integration, we multiply the slope of the probability distribution in Eq. \eqref{eq:Slope2} by the neutrino mass, $m$, and integrate it over $r$. The result, shown in Fig. \ref{fig:qua_fit}, is the mass distribution of the neutrino atom in the galaxy.  The integral of this distribution is $4.25\times10^{-18}\textup{M}_\odot$ which is $10^{-29}$ times smaller than the expected dark matter mass in the Milky-Way which is estimated to be $1.85\times10^{11}\textup{M}_\odot$  \cite{jiao2023}. Despite the mathematical elegance of the neutrino atom approach, our analysis led to its rejection as a candidate for dark matter as it is not massive enough to explain the rotational curve of the galaxy. 

Assuming that low energy neutrinos are not the main component of dark matter in the galaxy we determine their atom mass distribution in the presence of dark matter not made of neutrinos. This distribution, shown in Fig.  \ref{fig:qua2_fit}, is slightly more massive than the previous one. Nevertheless, if the distribution in Fig.  \ref{fig:qua2_fit} could be detected in our galaxy it would be direct evidence of the existence of the graviton \cite{flores_qed}.

If gravity is just a deformation of spacetime as assumed in general relativity, then neutrinos do not experience quantum interactions outside the range of $10^{-18}m$  \cite{flores_qed}. Thus, in the galactic environment, low energy neutrinos are only bound by the conservation laws and they behave as classical particles. To test this hypothesis, we model the halo as a fluid that obeys Euler's equation and an equation of state that relates density, pressure, and $v_{rms}$ speed.

Euler's equation for non-viscous fluid at rest is given by
\begin{equation}
    \frac{1}{\rho} \frac{d}{dr}(P) + g(r) = 0
    \label{eq:Euler_eq}
\end{equation}
where $\rho$ is the neutrino density, $P$ is pressure and $g$ is the total gravitation field.  The unit of mass is $10^9\textup{M}_\odot$ and the unit of distance is $kpc$. The equation of state is the ideal gas law but with the temperature replaced by the root mean square velocity, $kT/m\hookrightarrow v^2_{rms}/3$, which results in a equation of state given by
\begin{equation}
    P = \frac{1}{3} v_{rms}^2 \rho.
\end{equation}
The reason for avoiding the use of temperature is that, near the galactic center, a neutrino could have a large outward velocity associated with relatively high temperature. Such a neutrino could easily travel to the edge of the galaxy without interacting with any other particle; at that point, its speed would be nearly zero, corresponding to a much lower temperature. For this reason, we find it more natural to use the root-mean-square speed, $v_{rms}(r)$,  rather than temperature in the ideal gas law.

The total gravitational field $g(r)$ in Euler's equations is:
\begin{equation}
    g(r) = g_{halo}(r) + g_{bulge}(r) + g_{disk}(r)
    \label{eq:total_g_field}
\end{equation}
where $g_{halo}$ is the self-gravitating field of the neutrinos within radius $r$. We assume that the central mass of the galaxy (the bulge) has a spherically symmetric density, $\rho_B$; the corresponding gravitational field is $g_{bulge}$. In contrast, the disk containing stars, gas, and dust does not possess spherical symmetry. Therefore, we separate the gravitational contribution of the disk and evaluate its effect on the neutrino density, $\rho$, both along the plane of the disk and along the disk’s axis of symmetry. We then compute a weighted average density, $\rho$. Our calculations below show that the disk does not significantly disrupt the spherical symmetry of the neutrino halo.

\begin{figure}[H]
    \centering
    \begin{subfigure}[b]{0.48\textwidth}
    \centering
\includegraphics[width=\textwidth]{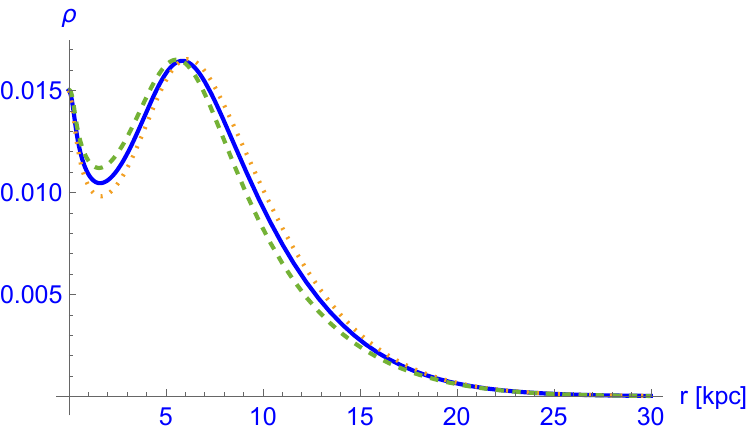}
    \end{subfigure}
    \begin{subfigure}[b]{0.48\textwidth}
    \end{subfigure}
    \caption{\textbf{Neutrino density as a function of radial distance.} The dashed green line is the density calculated along the galactic disk, the dotted orange line is the density calculated along the axis of symmetry of the galaxy and the solid blue line is the weighted average of the two densities. We note that the non-spherically symmetric galactic disk does not seem to affect significantly the density distribution. According to this plot, the neutrino density at $8.3 kpc$, the location of the solar system, is $0.0136\times10^9\textup{M}_\odot/kpc^3$. }
    \label{fig:density_fit}
\end{figure}

Substituting the different components into Euler's equation yields:
\begin{equation}
    \frac{1}{3} \frac{r^2}{\rho} \frac{d}{dr}(v_{rms}^2 \rho) + r^2 g_{disk}(r) + 4\pi G \int_{r_0}^{r} t^2 [\rho(t) + \rho_B(t)] dt = 0
    \label{eq:density_eq}
\end{equation}
The distribution of baryonic mass that results in $\rho_B$  and $g_{disk}$ is similar to the one used by Jiao et al. \cite{jiao2023}. Solving Eq. \eqref{eq:density_eq} yields the graph in Fig. \ref{fig:density_fit}. The curve with the dashed green line is the neutrino density calculated along the galactic disk. The curve with the dotted orange line is the neutrino density along the axis of symmetry perpendicular to the disk. The curve with the solid blue line is the weighted average of the two. 

\begin{figure}[H]
    \centering
    \begin{subfigure}[b]{0.48\textwidth}
        \includegraphics[width=\textwidth]{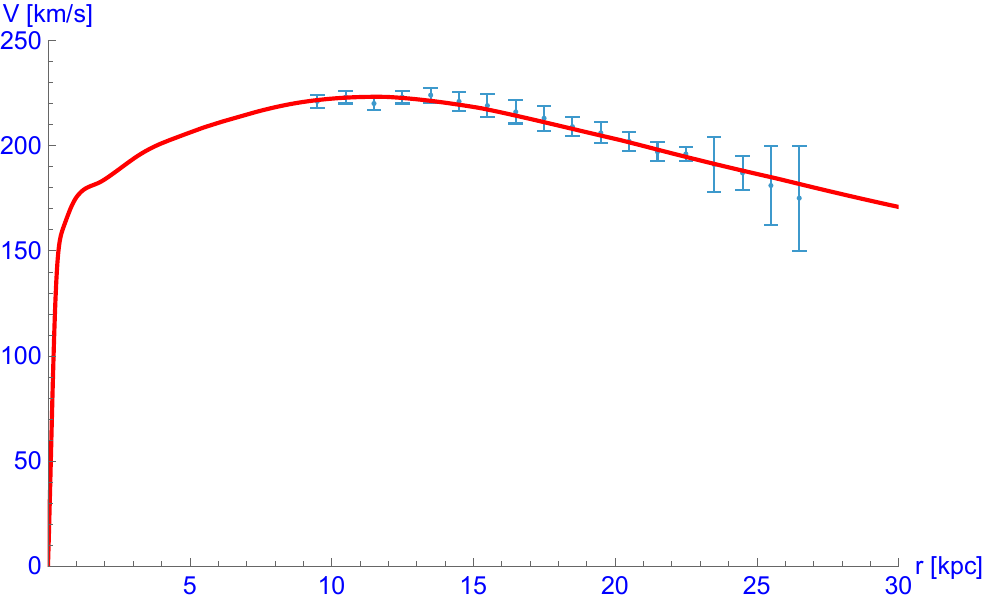}
        \caption{Best fit}
              \label{fig:rotational_fit}
    \end{subfigure}
    \hfill
    \begin{subfigure}[b]{0.48\textwidth}
        \includegraphics[width=\textwidth]{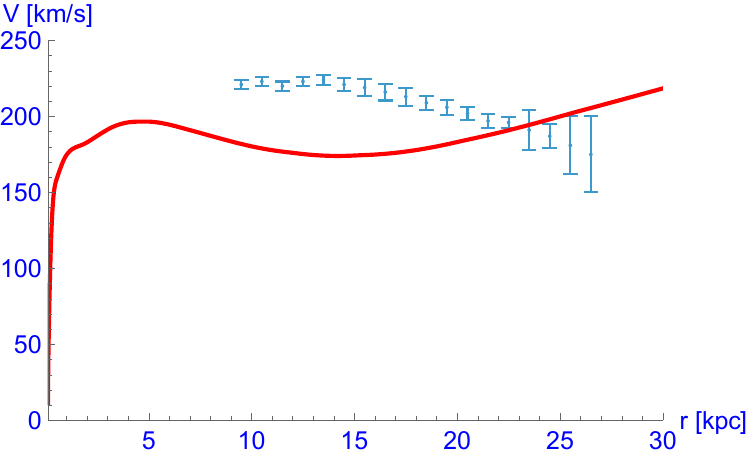}
        \caption{Isothermal halo}
        \label{fig:isothermal}
    \end{subfigure}

    \caption{\textbf{Rotational curve for the Milky Way galaxy.} (a) Assuming that $v_{rms}$ is a function of radius, we determine the form of this function that when inserted into Euler’s equation, provides an accurate fit to the observed rotational curve.
 (b) The isothermal classical fluid model fails to reproduce the galactic rotational curve for any choice of temperature.}
\end{figure}

Once the neutrino density, $\rho$, has been determine, we calculate the gravitational field, $g_{halo}(r)$. Now we can calculate the rotational curve along the galactic disk using the equation, $v=\sqrt{g(r)r}$, where $g$ is the net gravitational field along the galactic disk in Eq. \eqref{eq:total_g_field}. The rotational curve, similar to the one found by Jiao et al \cite{jiao2023}, for the Milky Way is plotted in Fig. \ref{fig:rotational_fit}. 

\begin{figure}[H]
    \centering
    \begin{subfigure}[b]{0.48\textwidth}
        \includegraphics[width=\textwidth]{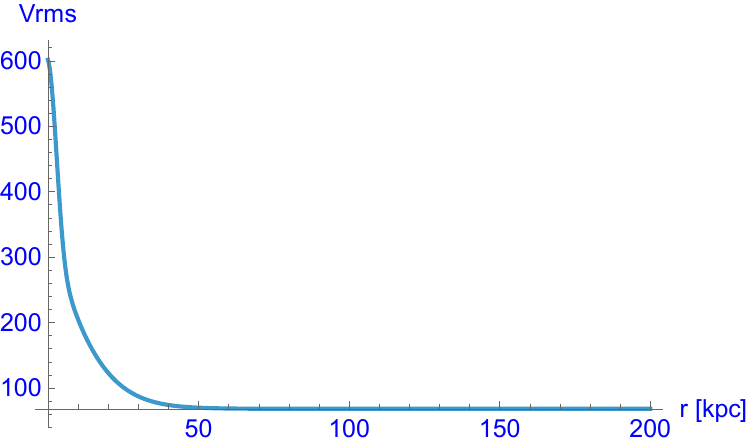}
        \caption{Neutrino $V_{rms}$ in the galaxy}
              \label{fig:Vrms}
    \end{subfigure}
    \hfill
    \begin{subfigure}[b]{0.48\textwidth}
        \includegraphics[width=\textwidth]{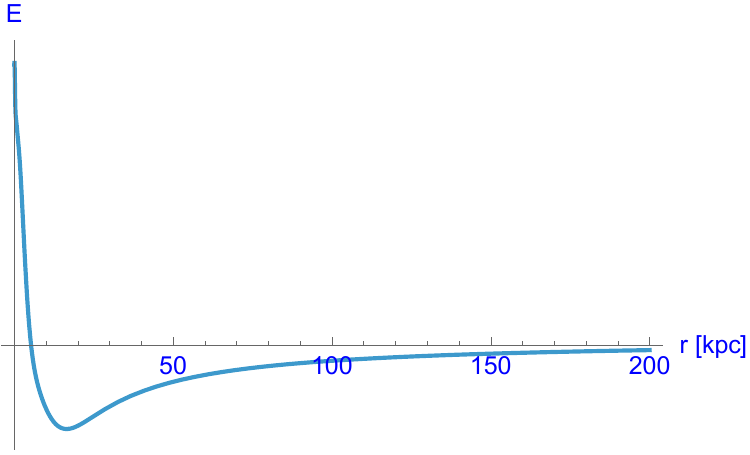}
        \caption{Average energy of a neutrino}
        \label{fig:energy_fit}
    \end{subfigure}

    \caption{\textbf{Neutrino $V_{rms}$ and average energy at different radii.} (a) Near the center of the galaxy, the $v_{rms}(r)$ speed changes quickly with radius. The velocity is in $km/s$. (b) The average energy of a neutrino is plotted as a function of radius up to a distance of $200kpc$. The energy is negative for most of the halo, which is physically consistent with a bound system.}

\end{figure}

The $v_{rms}$ profile shown in Fig.  \ref{fig:Vrms} is the one that reproduces a rotation curve consistent with the observations of Jiao \textit{et al.} This function is given by:
\begin{equation}
    V^2_{rms} =2.25\times10^5e^{-(r/3.4)^2}+1.32\times10^5e^{-r/7.9}+4.64\times10^3
    \label{eq:Vrms}
\end{equation}
From physical considerations, it is reasonable to expect that in regions where the gravitational potential is deeper $v_{rms}(r)$ with larger values of should appear. We consider the function $v_{rms}(r)$ as a prediction to be tested again experimental observation. The average neutrino energy is shown in Fig.  \ref{fig:energy_fit}. As expected for bound particles in orbit, the average energy is negative over most radii. 

Reionization-limited H I clouds, or RELHICs, are compact clouds of atomic hydrogen gas that are gravitationally confined by dark matter halos but lack sufficient mass to form galaxies. RELHICs are considered an important prediction of the $\Lambda$CDM model and may provide insight into the early universe \cite{Benitez2017}. One of the most promising RELHIC candidates discovered thus far is “Cloud-9,” first identified in 2023. Because Cloud 9 consists exclusively of hydrogen and dark matter, it is comparatively much simpler to model than large galactic systems such as the Milky Way.

The hydrogen gas in Cloud-9 is believed to be in thermal equilibrium with the ultraviolet background radiation field (UVB) and in hydrostatic equilibrium with the surrounding dark matter halo \cite{Gagandeep2025}.  These properties make Cloud-9 an ideal candidate for investigating simplified dark matter models under the assumption of spherical symmetry.

\begin{figure}[H]
    \centering
    \begin{subfigure}[b]{0.48\textwidth}
        \includegraphics[width=\textwidth]{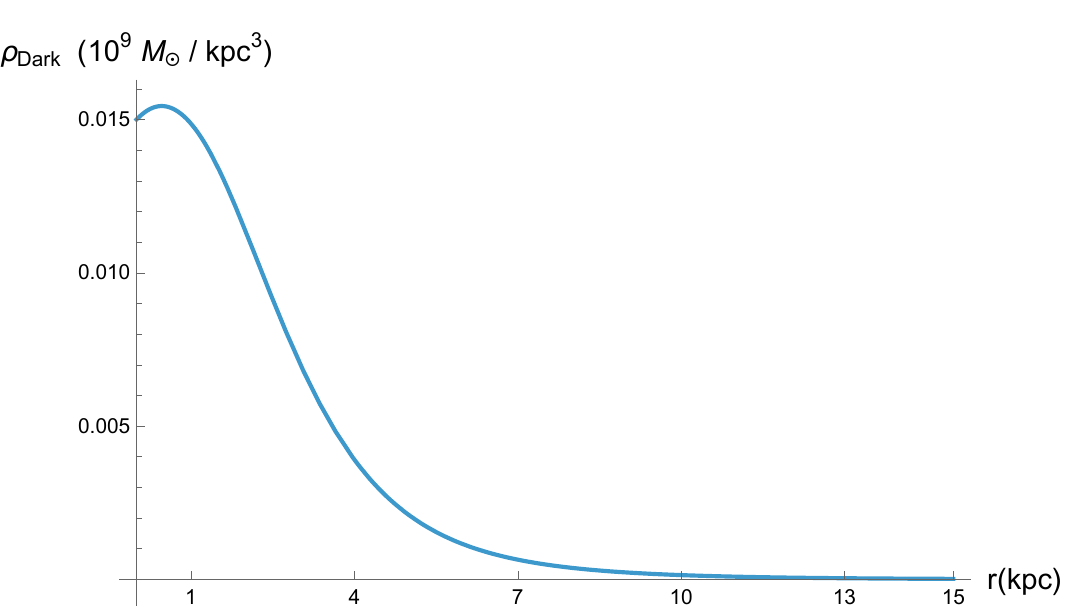}
        \caption{Dark matter density}
              \label{fig:Untitled 12}
    \end{subfigure}
    \hfill
    \begin{subfigure}[b]{0.48\textwidth}
        \includegraphics[width=\textwidth]{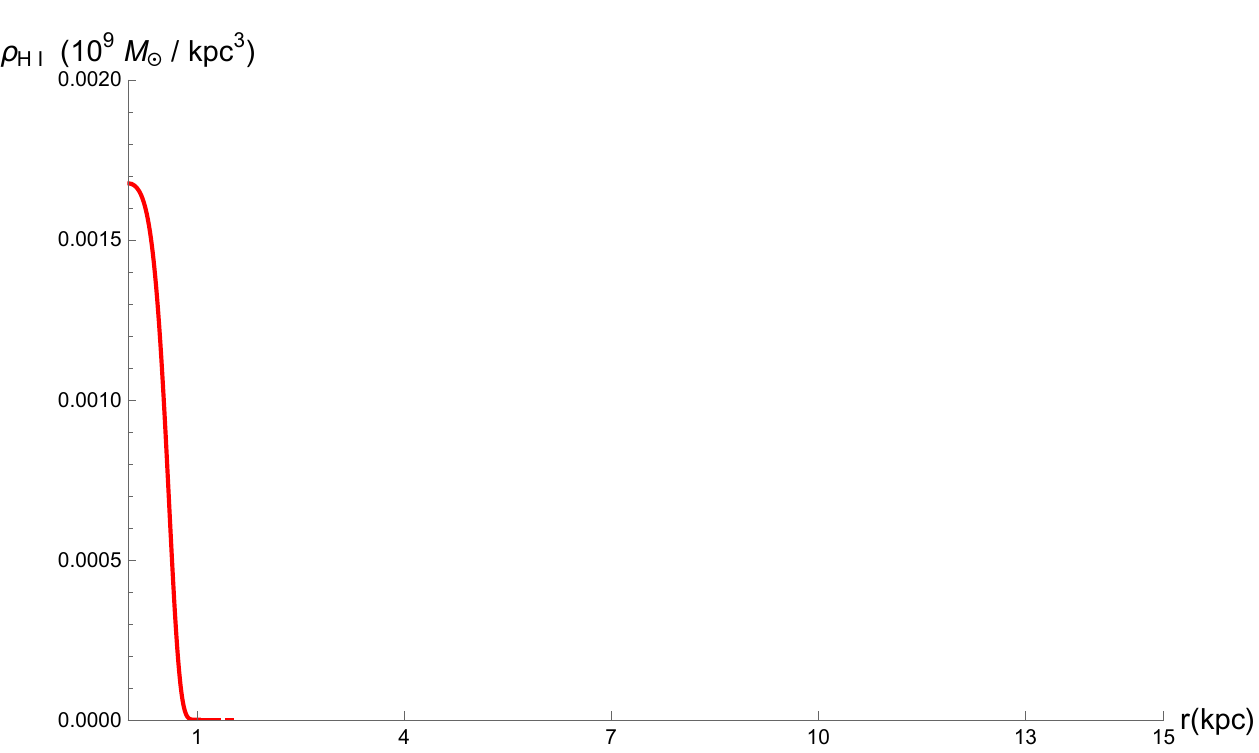}
        \caption{Hydrogen density}
        \label{fig:Untitled 13}
    \end{subfigure}

    \caption{\textbf{Dark matter and H I density for Cloud-9 as a function of radial distance.} (a) The dark matter halo extend as far as 12 kpc. The resulting mass and density distributions of dark matter align qualitatively with the extended halo structure predicted by the $\Lambda$CDM model. (b) The hydrogen cloud is confined primarily to the central bulge region and tapers off near a radius of approximately 1 kpc.}

\end{figure}

For the case of free, collisionless neutrinos, we apply the same mathematical technique used for the Milky Way, now adapted to full spherical symmetry. We begin with an assumed hydrogen distribution consistent with the observed values \cite{Gagandeep2025},  and use Euler’s equation to obtain the corresponding neutrino distribution. We then use Euler’s equation again to obtain the hydrogen distribution in the background of the previously obtained neutrino distribution. We iterate this process 11 times until the values inferred from observation of Cloud-9 are stable to three significant figures. The results of our calculation of the dark matter density shown in Fig.  \ref{fig:Untitled 12} and the hydrogen (H I) density shown in Fig.  \ref{fig:Untitled 13} .

In this work, we have explored the behavior of low-energy galactic neutrinos under two fundamentally different paradigms for gravity, and in doing so we have shown that neutrinos provide a uniquely sensitive probe of gravity’s true nature.

If gravity is a quantum force, its intrinsically long-range interaction necessarily gives rise to an atom-like bound neutrino structure. Although our calculations demonstrate that the resulting neutrino atom is far too light—by roughly $10^{-29}$—to account for dark matter, its significance is profound. Experimental confirmation of the structure shown in \ref{fig:qua2_fit} would constitute direct, model-independent evidence that gravity is a quantum interaction mediated by gravitons. In this sense, the neutrino atom serves not as a dark-matter candidate, but as a potential \textit{smoking-gun signature} of quantum gravity.

If, instead, gravity is not a quantum force but emerges solely from spacetime curvature, neutrino interactions are governed exclusively by the weak force, whose extremely short range ($10^{-18}m$) renders low-energy galactic neutrinos effectively collisionless. In this regime, low energy neutrinos behave as free classical gas orbiting the galaxy. Our results show that the classical neutrino gas cannot be isothermal. We predict, independently of the neutrino mass value, that the $v_{rms}(r)$ speed is a function of the radial distance to the center of the galaxy and we proposed that this function could be in principle experimentally verified. For instance, according to Eq. \eqref{eq:Vrms}, at the solar system, $r=8.3 kpc$, we find $v_{rms}=226.7 km/s$. At this location, the graph in Fig. \ref{fig:density_fit}. predicts that the neutrino density is $0.0136\times10^9\textup{M}_\odot/kpc^3$. Importantly, our results demonstrate that free classical neutrinos remain a viable dark-matter candidate in a non-quantum gravity framework.

Finally, in the absence of quantum gravity, large populations of low-energy neutrinos could remain gravitationally bound to galaxies. In this context, the possibility that a substantial fraction of these particles are antineutrinos becomes particularly intriguing, offering a potential explanation for the apparent scarcity of antimatter in the observable universe. Such a scenario would require lepton-number violation at very high energies, pointing naturally toward physics beyond the Standard Model.

Taken together, these results show that low-energy neutrinos sit at a critical intersection of cosmology, particle physics, and gravitation. Whether gravity is quantum or classical, neutrinos provide testable, falsifiable consequences that link galactic dynamics to fundamental physics. In this way, neutrinos emerge not only as potential dark-matter constituents, but also as powerful messengers capable of revealing the true nature of gravity itself.

\section*{Acknowledgment}

We would like to thank the students who in past semesters have taken part on this research project.


\begin{thebibliography}{99}

\bibitem{KATRIN2025}
KATRIN Collaboration, Science Vol. 338, Issue 6743 (2025)

\bibitem{degraaf1966}
T. De Graaf, H.A. Tolhoek, \textit{Nuclear Physics} 81, 3 (1966).

\bibitem{tremaine1979}
Tremaine, S., \& Gunn, J. E. (1979). Dynamical role of light neutral leptons in cosmology. \textit{Physical Review Letters}, 42(6), 407.

\bibitem{flores_qed}
Flores, E. V., Quantum Mechanics Interpreted Through Quantum Electrodynamics, arXiv:2208.12267.

\bibitem{bjorken1964}
J. D. Bjorken \& S. D. Drell, \textit{Relativistic Quantum Mechanics} (McGraw-Hill, New York, 1964).

\bibitem{Xu}
X. Xu and E. R. Siegel, arXiv:0806.3767v1

\bibitem{greiner2009}
W. Greiner and J. Reinhardt, {\textit{Quantum Electrodynamics}} (Springer, 2009), Fourth Edition.

\bibitem{Cotaescu}
Ion I. Cotaescu, (2007) {\textit{Mod.Phys.Lett.}}A22:2493-2498

\bibitem{LDMF}
Library of Digital Mathematical Functions, Eq. 13.2.6 https://dlmf.nist.gov/13.2

\bibitem{jiao2023}
Jiao, Y., et al. (2023). Detection of the Keplerian decline in the Milky Way rotation curve. \textit{{Astronomy \& Astrophysics}}

\bibitem{Benitez2017}
Benítez-Llambay, A. et al., The Properties of ‘Dark’ $\Lambda$CDM Haloes in the Local Group. {Monthly Notices of the Royal Astronomical Society}, vol. 465, no. 4, 11 Mar. 2017.

\bibitem{Gagandeep2025}
Gagandeep S. A. et al., The First RELHIC? Cloud-9 Is a Starless Gas Cloud. arXiv:2508.20157v2, 2025.

\end{thebibliography}
\end{document}